\documentclass[graphicx,floatfix,aps,pre,epsfig,showpacs]{revtex4}

\usepackage{graphicx}
\usepackage{amsmath}
\begin{document}

\title{Propagation of external regulation and asynchronous
dynamics\\ in random Boolean networks}

\author{H. Mahmoudi}
\affiliation{Institute for Scientific Interchange, Viale Settimio Severo 65, 
Villa Gualino, I-10133 Torino, Italy}
\affiliation{Dipartimento di Fisica Teorica, Universita degli Studi di
Torino, Via Giuria 1, I-10125 Torino, Italy}

\author{A. Pagnani}
\affiliation{Institute for Scientific Interchange, Viale Settimio Severo 65, 
Villa Gualino, I-10133 Torino, Italy}

\author{M. Weigt}
\affiliation{Institute for Scientific Interchange, Viale Settimio Severo 65, 
Villa Gualino, I-10133 Torino, Italy}

\author{R. Zecchina}
\affiliation{International Centre for Theoretical Physics, Strada
Costiera 11, P.O. Box 586, I-34100 Trieste, Italy}
\affiliation{Politecnico di Torino, C.so Duca degli Abruzzi 24, 
Torino, Italy}

\date{\today}

\begin{abstract}
Boolean Networks and their dynamics are of great interest as abstract 
modeling schemes in various disciplines, ranging from biology to computer 
science. Whereas parallel update schemes have been studied extensively in 
past years, the level of understanding of asynchronous updates schemes is 
still very poor. In this paper we study the propagation of external 
information given by regulatory input variables into a random Boolean 
network. We compute both analytically and numerically the time evolution 
and the asymptotic behavior of this propagation of external regulation 
(PER). In particular, this allows us to identify variables which are
completely determined by this external information. All those variables 
in the network which are not directly fixed by PER form a core which 
contains in particular all non-trivial feedback loops. We design a 
message-passing approach allowing to characterize the statistical
properties of these cores in dependence of the Boolean network and the
external condition. At the end we establish a link between PER dynamics 
and the full random asynchronous dynamics of a Boolean network.
\end{abstract}


\maketitle 

\textbf{ The main motivation for this work is to study the propagation of
external information given by regulating but non-regulated variables into
random Boolean networks. This process, called {\it propagation of external
regulation}, eventually stops due to one of two reasons: Either the external
information has propagated throughout the full network, or a core of
variables cannot be fixed by PER. The statistical properties of these cores
are determined by the ratio $\alpha$ between the numbers of Boolean 
constraints and of variables, and by the composition of the constraints.
In this work we will {\em embed} the propagation dynamics of the
external condition into the asynchronous update dynamics by
introducing ternary variables of values $ \{ 0,1 ,\star \}$. 
The supplementary {\em joker} value $\star$ indicates that a variable
has not been fixed to a Boolean value in the PER dynamics. The introduction 
of this new value $\star$ allow us to analyze the PER dynamics in terms of 
a new constraint satisfaction problem with the same topology of the original
one, but where Boolean constraints are extended in a natural way in
order to include this ternary representation. This observation allows us
to apply recently developed tools from the statistical-mechanics approach
to combinatorial optimization problems.}

\section{Introduction}

Boolean Networks (BNs) are dynamical models originally introduced by
S. Kauffman in the late 60s \cite{Kauf69}. Since Kauffman's seminal
work, they have been used as abstract modeling schemes in many
different fields, including cell differentiation, immune response,
evolution, and gene-regulatory networks (for an introductory review
see \cite{GCK} and references therein). In recent days, BNs have
received a renewed attention as a powerful scheme of data analysis and
modeling of high-throughput genomic and proteomic experiments
\cite{IlyaBook2005}. Considerable attention has been given in the
previous literature to the classification of different attractor types
present in BNs under deterministic parallel update dynamics
\cite{Kauf69,Kauf1,samuelsson,BastollaParisi}. A special relevance has
been attributed to the so-called critical BNs \cite{Derrida} situated
at the transition between an ordered and a chaotic regime.

The original dynamical problem can been cast into a {\em constraint
satisfaction problem} \cite{MPZ,Libro_Martin}. Following
this approach presented in \cite{letter,jstat,fsc-jstat} and partially
also in \cite{Lagomarsino}, it is possible to study the organization
of fixed points in the thermodynamic limit in random Boolean
networks. This leads to the identification of a transition characterized 
by the sudden emergence of a computational core. Its existence is 
necessary but not sufficient for a globally complex phase to exist where 
fixed points are organized in an exponential number of macroscopically
separated clusters. This phenomenon is found to be robust with respect
to the choice of the Boolean functions. It is missing only in networks
where all Boolean functions are of {\sf AND} or {\sf OR} type. The size 
of the complex regulatory phase is found to
grow with the number $K$ of inputs of the Boolean functions.

The organization of the paper is the following: in
Sec.~\ref{sec:model} we introduce the model focusing to Kauffman
models with $K=2$ inputs per Boolean function, in Sec.~\ref{sec:per}
we introduce the notion of {\em propagation of external regulation},
the message-passing algorithm we will use for analyzing the problem.
We also discuss in details the resulting phase diagram of the
model. In Sec.~\ref{sec:async_dyn} we will give some analytical
prediction on the random asynchronous dynamics of random Boolean
networks, and its relation to the PER dynamics. Conclusions are drawn
in Sec.~\ref{sec:concl}.

\section{The model}
\label{sec:model}

\begin{figure}[htb]
\vspace{0.2cm}
\begin{center}
\includegraphics[width=0.75\columnwidth]{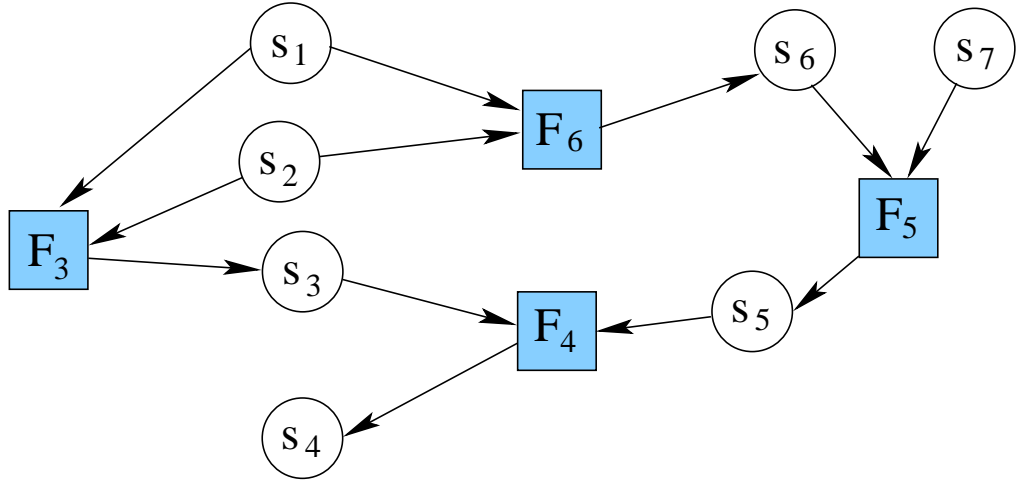}
\end{center}
\caption{Factor graph representation of a small Boolean network:
circles symbolize variables, squares Boolean functions. $s_1$
is an example for an external input variable, $s_4$ for a functional
variable, whereas $s_3$ stands for a regulatory variable.}
\label{fig:fig1}
\end{figure}

Let us first define properly the model we are going to investigate. It
is formed by $N$ Boolean variables collected in a vector $\vec s =
(s_1, \dots, s_N) \in \{0,1\}^N$; in Fig.~\ref{fig:fig1} they are
represented by circles. They are constrained by Boolean functions
$F_a(s_{a_1},s_{a_2})$ with $a\in INT \subset \{1,...,N\}$ running
over $M=|INT|$ different variables, each depending on two input
variables. The generalization to more than two inputs is obvious, but
in this work we will concentrate fully on the two-input case. These
functions are represented by squares in Fig.~\ref{fig:fig1} within the
so-called {\it factor-graph representation} of a BN.

The functions $F_a$ allow to define a {\it random asynchronous update
dynamics}, cf.~\cite{Gersh,Drossel}: In each step an
element $a$ of $INT$ is selected randomly and updated according to
its regulating Boolean function
\begin{equation}
\label{eq:function}
s_a^{T+1} = F_a(s_{a_1}^T,s_{a_K}^T)\ ,
\end{equation}
with ${\vec s}^{~T}$ denoting the configuration after $T$ time steps. This
time step is repeated, after $M$ updates every function is selected on
average once.

In this work, we concentrate our attention on {\it random factor
graphs} subject to two conditions:
\begin{itemize}
\item[({\em a})] Function nodes $F_a$ have fixed in-degree $2$ and
out-degree one ({\it i.e.}, two input and one output variable).
\item[({\em b})] Variables $s_a$ have in-degree at most one, {\it
i.e.}, either they depend on one single function, or they are not
regulated by any function.
\end{itemize}
Setting $\alpha := M/N$, the degree distribution of variable nodes
approaches asymptotically
\begin{eqnarray}
\rho^{\mathrm{out}}(d_{\mathrm{out}}) &=& e^{-K \alpha }
\frac{(K\alpha)^{d_{\mathrm{out}}}}{d_{\mathrm{out}}!} \nonumber \\
\rho^{\mathrm{in}}(d_{\mathrm{in}}) &=& \alpha \delta_{d_{\mathrm{in}},1}
+ (1-\alpha)\delta_{d_{\mathrm{in}},0}
\label{eq:degdistr}
\end{eqnarray}
{\em i.e.}, the out-degree distribution is a Poissonian of mean
$K\alpha$, while in-degree distribution is bimodal. Since in- and
out-degrees are uncorrelated, the joint degree distribution
factorizes: $\rho(d_{\mathrm{out}},d_{\mathrm{in}}) =
\rho^{\mathrm{out}}(d_{\mathrm{out}})
\rho^{\mathrm{in}}(d_{\mathrm{in}})$. Random factor graphs are
obviously a drastic oversimplification of realistic models of
gene-regulatory networks: There available data show evidence for a 
broad, possibly scale-free $\rho^{\rm out}(d_{\rm out})$ and a more 
concentrated in-degree 
distribution being compatible with an exponential form
\cite{GuBoBoKe,Albert}. Nevertheless, this simplified model allows for
many detailed analytic predictions that can guide our comprehension in
more realistic and interesting cases, and is a well-controlled testing
ground for techniques borrowed from statistical mechanics. Such
techniques are not limited to random graphs and can be easily extended
to deal with more realistic cases.
\begin{figure}[htb]
\vspace{0.2cm}
\begin{center}
\includegraphics[width=0.75\columnwidth]{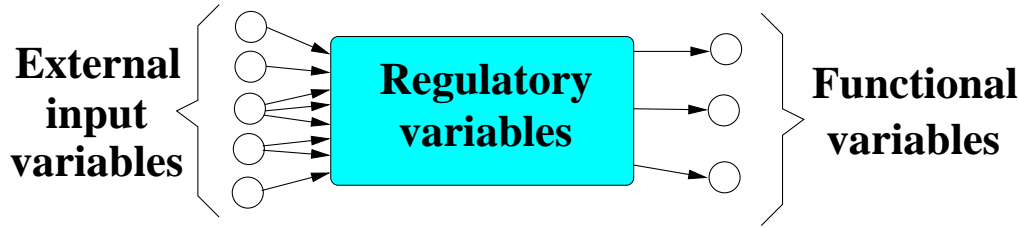}
\end{center}
\caption{Schematic representation of the three types of variables:
external input, functional, and regulatory.}
\label{fig:sandwich}
\end{figure}

In general, we can distinguish three sets of variables as
displayed in Fig.~\ref{fig:sandwich}:
\begin{itemize}
\item {\it External input variables:} There are $N-M = (1-\alpha)N$
variables which are not regulated by any function, they are collected
in the set $EXT=\{1,..,N\}\setminus INT$. They represent external inputs to
the network. We do not consider these external inputs as dynamical
variables of the system. In the original definition of Kauffman
networks they are in fact included into the definition of the BN
itself: This definition works at $\alpha=1$ but allows for constant
functions, whose outputs are the analogues of our external input
variables.
\item {\it Regulatory variables} are all those variables which are
regulated and regulate.
\item {\it Functional variables:} There are $e^{-2\alpha} N$
variables which do not regulate any other function. 
\end{itemize}
The last both types form together the set $INT$ of internal, regulated
variables. One has to note that in a random network there are some
variables which belong theoretically to both the external and the
functional variables because they are isolated. They contribute only
trivially to the behavior of the network and can therefore be
neglected in the discussion.

\begin{table}[h]
\begin{tabular}{|cc|cc|cccc|cccccccc|cc|}
\hline $s_1$ &
$s_2$&0&1&$s_1$&$\overline{s_1}$&$s_2$&$\overline{s_2}$
&$\wedge$&&&&&&&$\vee$&$\oplus$&$\overline{\oplus}$\\
\hline 0 & 0 & 0 & 1 & 0 & 1 & 0 & 1 & 0 & 0 & 0 & 1 & 1 & 1 & 1 & 0 &
0 & 1\\ 0 & 1 & 0 & 1 & 0 & 1 & 1 & 0 & 0 & 0 & 1 & 0 & 1 & 1 & 0 & 1
& 1 & 0\\ 1 & 0 & 0 & 1 & 1 & 0 & 0 & 1 & 0 & 1 & 0 & 0 & 1 & 0 & 1 &
1 & 1 & 0\\ 1 & 1 & 0 & 1 & 1 & 0 & 1 & 0 & 1 & 0 & 0 & 0 & 0 & 1 & 1
& 1 & 0 & 1\\ \hline
\end{tabular}
\caption{Truth table for all 16 boolean functions of $K=2$ inputs.}
\label{tab:k2fun}
\end{table}

We have to specify the functions acting on top of the random
topology defined so far.  There are $2^{2^K}=16$ Boolean functions,
which can be grouped into 4 classes \cite{Kauf2}:
\begin{itemize}
\item[(i)] The two constant functions. 
\item[(ii)] Four functions depending only on one of the two
inputs, {\em i.e.} $s_1,\overline s_1, s_2,\overline s_2$. 
\item[(iii)] {\it Canalizing functions (AND-OR class):} There are
eight functions, which are given by the logical AND or OR of the two
input variables, or of their negations. These functions are {\it
canalizing}. If, {\it e.g.}, in the case $F(s_1,s_2)=s_1 \wedge s_2$
the value of $s_1$ is set to zero, the output is fixed to zero
independently of the value of $s_2$.  It is said that $s_1$ is a {\it
canalizing variable} of $F$ with the {\it canalizing value} zero.
\item[(iv)] {\it Non-canalizing functions (XOR class):} The last two
functions are the XOR of the two inputs, and its negation. These two
functions are not canalizing, whatever input is changed, the output
changes too.
\end{itemize}

We keep in mind that the case of Boolean functions depending on
exactly $K=2$ input variables is just the simplest interesting case.
Since the number of Boolean functions of $K$ variables increases as
$2^{2^K}$, a complete classification of Boolean function becomes
intractable already for relatively small $K$.  For clarity we
therefore concentrate first on true $(K=2)$-functions only, {\em
i.e.}~on those in the canalizing AND-OR class and the non-canalizing
XOR class. We therefore require $x M$ functions to be in the XOR
class, and the remaining $(1-x)M$ functions to be of the AND-OR type,
with $0\leq x\leq 1$ being a free model parameter. In this sense, for
$K=2$, the network ensemble is completely defined by $\alpha$ and
$x$. It is interesting to note that the case $x=1$ on a slightly
different class of random hypergraphs, has been already studied in a
different context in \cite{XOR}.

\section{Propagation of external regulation}
\label{sec:per}
\begin{figure}[htb]
\vspace{0.2cm}
\begin{center}
\includegraphics[width=0.75\columnwidth]{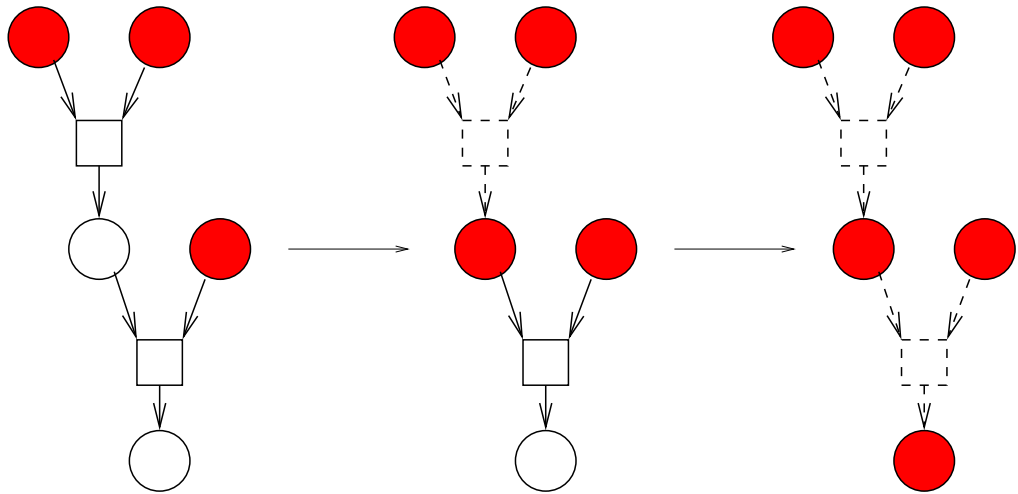}
\end{center}
\caption{PER: Both inputs to the upper left Boolean function are
  external, so also the output is directly fixed. Once this is done,
  also the inputs to the second function are fixed, and again the
  information can be propagated. All variables in this small sample
  graph are therefore determined by PER, no core exists.}
\label{fig:per}
\end{figure}

An important dynamical process in BN is the propagation of the
external condition given by the non-regulated variables into the
network. Imagine, {\it e.g.}, that all inputs of a Boolean
function are external variables. Fixing the external condition, also
the output of the considered Boolean function is fixed, the external
condition has been propagated. Iterating this propagation, we may
eventually also fix variables which do not depend directly on external
variables, but whose inputs have been fixed in an earlier iteration
step, cf. Fig.~\ref{fig:per}. Note that in the case of canalizing
functions it is sufficient to have one input variable fixed to its
canalizing value in order to be able to propagate the information.
Thus, PER is more efficient in the case of a small fraction $x$
of non-canalizing, XOR-type functions in the network.

This propagation of external regulation may stop in two ways: First,
all variables might be fixed, and the external information is
propagated throughout the full network. Second, a core of functions
survives which still have unfixed output. This {\it PER core} contains
in particular all relevant feed-back loops which are not broken due to
the inclusion of canalizing functions.

The propagation {\it dynamics} of the external condition can be
embedded into the asynchronous update dynamics of
Eq.~(\ref{eq:function}) by generalizing the Boolean variables to
ternary variables, and the Boolean functions accordingly. In
addition to the values 0 and 1 we introduce the {\it joker value}
$\star$ for variables having no fixed value. As an initial condition
for PER, only the external variables have assigned Boolean values,
whereas all nodes belonging to $INT$ are assigned value
$\star$. Generalizing the functions according to
Tab.~\ref{tab:k2fun_ext}, the dynamics of PER is given exactly by
Eq.~(\ref{eq:function}), with a randomly selected $a\in INT$ for each
single-variable update.
 
\begin{table}[h]
\begin{tabular}{|cc|cccccccc|cc|}
\hline $s_1$ & $s_2$ &$\wedge$&&&&&&&$\vee$&$\oplus$&$\overline{\oplus}$\\
\hline 
0 & 0 & 0 & 0 & 0 & 1 & 1 & 1 & 1 & 0 & 0 & 1\\ 
0 & 1 & 0 & 0 & 1 & 0 & 1 & 1 & 0 & 1 & 1 & 0\\ 
0 & $\star$ & 0 & 0 & $\star$ & $\star$ & 1 & 1 & $\star$ & $\star$ & $\star$ & $\star$\\
1 & 0 & 0 & 1 & 0 & 0 & 1 & 0 & 1 & 1 & 1 & 0\\ 
1 & 1 & 1 & 0 & 0 & 0 & 0 & 1 & 1 & 1 & 0 & 1\\
1 & $\star$ & $\star$ & $\star$ & 0 & 0 & $\star$ & $\star$ & 1 & 1 & $\star$ & $\star$\\
$\star$ & 0 & 0 & $\star$ & 0 & $\star$ & 1 & $\star$ & 1 & $\star$ & $\star$ & $\star$\\ 
$\star$ & 1 & $\star$ & 0 & $\star$ & 0 & $\star$ & 1 & $\star$ & 1 & $\star$ & $\star$\\
$\star$ & $\star$ & $\star$ & $\star$ & $\star$ & $\star$ & $\star$ & $\star$ & $\star$ 
& $\star$ & $\star$ & $\star$\\
\hline
\end{tabular}
\caption{Extended truth table for canalizing and non-canalizing 
  functions of $K=2$ inputs.}
\label{tab:k2fun_ext}
\end{table}
                                                                               
Note that, the fraction of variables reached by PER depends not only
on the network but also on the precise external condition. In a
previous publication \cite{jstat}, we have analyzed the average case
which is realized by the vast majority of external conditions. Here
we are aiming at a more precise description of fluctuations of the PER
core size due to various external conditions. Note that such
fluctuations are completely due to the existence of canalizing
functions which may or may not propagate the information of a single
fixed input, in dependence on whether it has its canalizing value or
not. For $x = 1$ we thus expect no such fluctuations to exist, and the
PER core is completely topological in the sense that it depends only
on the underlying network and not the actual realization of the
functions.

\subsection{Belief propagation for the ensemble of PER cores}

The algorithmic determination of the PER core for a single external 
condition can be achieved in linear time  by simply following the
definition of the process itself. If we want to characterize the
statistical properties of the ensemble of {\it all} cores resulting
from $2^{(1-\alpha)N}$ external conditions, running times obviously
become exponential and thus not feasible for large networks.

The introduction of the joker value allows us to set up a directed
belief-propagation algorithm \cite{Yedidia, SumProd, BMZ} which may be
used to {\it efficiently} describe the statistical properties of all
regulated variables over all external conditions. Some interesting
limiting cases will be discussed separately: The appearance of the
first PER cores in an exponentially small fractions of all external
conditions, the appearance of the PER core for a typical external
condition, and the appearance of a non-zero overlap of all PER cores.

As already mentioned, we are interested in the behavior for the full
ensemble of external configurations. We therefore introduce the
single-site distributions
\begin{equation}
p_i(s) = 2^{-(1-\alpha)N} \sum_{\rm ext.\ cond.} \delta_{s_i,s}
\end{equation}
as the histogram of the value of variable $s_i$ over all external
conditions, after completion of PER. In this notation, an external
variable obviously has
\begin{equation}
p_i(s) = \frac 12 \delta_{s,0} + \frac 12 \delta_{s,1}, \ \ \ \ 
i\in EXT
\end{equation}
since by definition it never takes value $\star$, whereas internal
variable might have a non-trivial weight also in $p_i(\star)$.

Under PER, these probabilities get updated as
\begin{equation}
  \label{eq:bp}
  p_i(s_i) = \sum_{s_j,s_k} \delta_{s_i,F_i(s_j,s_k)}\ p_j(s_j)\ p_k(s_k)\ ,
\end{equation}
following the output of the extended Boolean function summed over all
possible input configurations. In this equation, variable $s_i$ is
understood to be regulated by $s_j$ and $s_k$. These equations can be
understood as a {\it directed Belief propagation} (BP). Note that the
BP equations contain a very important assumption: The probability of
the two regulating variables is factorized, {\it i.e.}, they are
assumed to be statistically independent. This assumption is expected
to become asymptotically exact in the thermodynamic limit $N\to\infty$
of infinitely large random Boolean networks. In finite networks, it 
might be violated
due to short loops: The input variables $s_i$ and $s_j$ might, {\it
e.g.}, depend on two Boolean functions with a common input and thus be
correlated. More generally correlations between two input variables are 
introduced whenever they have some common ancestor. In random network
such ancestors are known to typically have a distance of ${\cal O}(\ln 
N)$, and thus become irrelevant in the case of very large BNs (inside
on thermodynamic state).

The update can be achieved iteratively, starting from the initial
condition
\begin{equation}
  \label{eq:ic}
  p_i^0 (s) = \left\{
  \begin{array}{ccl}
    \frac 12 \delta_{s,0} + \frac 12 \delta_{s,1} && {\rm for}\ i\in EXT \\
    \delta_{s,\star} && {\rm for}\ i\in INT \\
  \end{array}
  \right.
\end{equation}
which reflects the fact that initially only the external variables are
fixed, and the regulated ones are still unknown.

Note that the original PER is included in Eq.~(\ref{eq:bp}) if the
external condition is polarized completely to a single configuration. In 
this moment, all marginal probabilities remain polarized into one of the
three possible values, and factorization in Eq.~(\ref{eq:bp}) becomes
trivially fulfilled. Under this initial condition, BP would be exact on 
whatever Boolean network.

Since BP equations are defined on a single graph,  
their applicability is not necessarily restricted to random BNs
with their Poissonian outdegree distribution. In the latter case one
can, however, introduce a simple average over the graph ensemble by
introducing the histogram of the $p_i(\cdot)$ for all internal
variables:
\begin{equation}
{\cal P}[\, p(\cdot)\, ] = \frac 1{\alpha N} \sum_{i\in INT}
\delta[\, p(\cdot)-p_i(\cdot)\,  ]
\end{equation}
with $\delta[\cdot]$ being a three-dimensional Dirac distribution in
all components of $p_i$. For a randomly chosen regulated variable,
each of the inputs is regulated with probability $\alpha$, and
external else. So we can write an effective self-consistent equation
for $P$,
\begin{eqnarray}
\label{eq:popdyn}
{\cal P}[\, p(\cdot)\, ] &=& 
\int {\cal D} p_1 {\cal D} p_2 
{\cal  Q}[\, p_1(\cdot), \alpha \,] 
{\cal  Q}[\, p_2(\cdot), \alpha \,] 
\nonumber \\ &\times&\left\langle \delta\left[ p(\cdot)- 
\sum_{s_1,s_2} \delta_{\cdot,F(s_1,s_2)}\ p_1(s_1)\ p_2(s_2)
\right]\right\rangle_F\,\,\,\,\,\,\,\,
\end{eqnarray}
where 
\begin{equation}
{ \cal Q}[\,p(\cdot), \alpha \,] = \alpha {\cal P}[\,
p(\cdot)\, ] + (1-\alpha) \delta\left[\, p_1(\cdot)
- \frac 12 (\delta_{\cdot,0} +  \delta_{\cdot,1}) \, \right]
\end{equation}
In this equation, each integration runs over a three-dimensional
simplex, and the average $\langle\cdot\rangle_F$ is taken over the
distribution of Boolean functions at given fraction $x$ of
non-canalizing functions. It can be solved iteratively by standard
methods as population dynamics, but some limiting cases can be
discussed analytically.

\subsection{The appearance of the first PER core}

For small $\alpha$, only few regulated variables exist in comparison
to the number of external variables. As already shown in
\cite{letter, jstat}, for $\alpha<1/(1+x)$ {\it typically} all variables
can be fixed by PER starting from the external condition. Typically
here means that, starting from a randomly chosen external
configuration and with probability approaching one in the thermodynamic
limit, no extensive PER core remains.

On the other hand, one should expect that an exponentially small
fraction of external
conditions already leads to a non-zero PER core at smaller $\alpha$.
This happens if external variables are chosen to take as rarely as 
possible canalizing values of the Boolean functions depending on them.

The argument can be made mathematically more stringent. For doing so,
we first observe that the existence of a PER core for some external
configurations leads to a finite fraction of variables $i$ with
$p_i(\star) > 0$. 

We therefore introduce the fractions of regulated variables never (resp.
always) taking a certain value $s$ out of $\{0,\star,1\}$
\begin{eqnarray}
\label{eq:tau_pi}
\tau_s&=& {\cal P}[ p(s)\equiv 0 ]\ =\ \frac 1{\alpha N} \sum_{i\in INT}
\delta_{p_i(s),0}\ , \nonumber\\
\pi_s&=& {\cal P}[ p(s)\equiv 1 ]\ =\ \frac 1{\alpha N} \sum_{i\in INT}
\delta_{p_i(s),1}\ .
\end{eqnarray}
The existence of an extensive core for some of external conditions
is obviously equivalent to $\tau_\star < 1$. Note that in our
unbiased ensemble of Boolean functions none of the values zero or one
is favored. We thus expect $\tau_0 = \tau_1$ and $\pi_0 = \pi_1$ for 
symmetry reasons.

\subsubsection{Fixed-point equations}

We can use the BP equations to get a finite closed set of equations
including $\tau_\star$. Technically this means to project
Eq.~(\ref{eq:popdyn}), which is formulated as an equation for
functions over a three-dimensional simplex, to a finite number of
variables. To achieve this, we have to discuss carefully all those
cases on the right-hand side of Eq.~(\ref{eq:popdyn}) which lead to a
vanishing probability for $s=\star$ on the left-hand side.
\begin{itemize}
\item For {\it non-canalizing} functions, a non-zero probability of a
$\star$-valued output exists whenever one of the inputs is allowed to
assume also value $\star$. To contribute to $\tau_\star$, both inputs
have to be non-$\star$, which happens with probability
$(1-\alpha+\alpha \tau_\star)^2$.
\item For {\it canalizing} functions, the situation is a bit more
involved. The output is prevented from taking value $\star$ also if
one of the inputs is frozen completely to its canalizing value, which
(due to the symmetry discussed above) happens with probability $\alpha
\pi_0=\alpha \pi_1$. To avoid double counting with the previously
discussed case, the second input than has to have a non-zero
probability in $\star$, which happens with $\alpha(1-\tau_\star)$
\end{itemize}
Putting the two cases together, we find 
\begin{eqnarray}
\label{eq:tau_star0}
\tau_\star &=& x(1-\alpha+\alpha \tau_\star)^2 +  (1-x)[ 
(1-\alpha+\alpha \tau_\star)^2 + 2 \alpha^2
\pi_0 (1-\tau_\star)] \nonumber\\
&=& (1-\alpha+\alpha \tau_\star)^2 + 2 (1-x) \alpha^2
\pi_0 (1-\tau_\star)\ .
\end{eqnarray}
This equation still depends on the probability $\pi_0=\pi_1$ that a
variable is frozen to one of the values 0 or 1. By definition, this
can happen only to regulated variables, but even there only if either
both inputs are frozen to a fixed value from $\{0,1\}$ or if one input
of a canalizing function is frozen to its canalizing value. In this
sense, frozen variables are generated only by other frozen
variables. On the other hand, the initial condition of
Eq.~(\ref{eq:ic}) does not contain such frozen values, and under
iteration of Eq.~(\ref{eq:bp}) they cannot be generated
spontaneously. We thus conclude
\begin{equation}
\pi_0=\pi_1=0\ ,
\end{equation}
and Eq.~(\ref{eq:tau_star0}) reduces to
\begin{equation}
\label{eq:tau_star1}
\tau_\star = (1-\alpha+\alpha \tau_\star)^2 \ .
\end{equation}
This equation does not depend on the fraction $x$ of canalizing
functions any more. It has two solutions, a trivial one relevant for
small $\alpha$, a non-trivial one for higher $\alpha$. More precisely
we find
\begin{equation}
\label{eq:tau_star}
\tau_\star = \left\{
\begin{array}{cl}
1 & {\rm if}\ \alpha<\frac 12 \\
\frac{(1-\alpha)^2}{\alpha^2} & {\rm if}\ \alpha>\frac 12 
\end{array}
\right.
\end{equation}
Note that also the solution $\tau_\star=1$ is physically sensible
beyond $\alpha = 1$. The derivation of Eq.~(\ref{eq:tau_star0})
requires only consistency of the generalized Boolean functions, which
is obviously also fulfilled for any fixed point of the BN containing no
$\star$-valued variables. The correct solution for the PER core is
given by both the consistency of functions and the maximality of the
number of $\star$, which results from the fact that only those
internal variables are changed from $\star$ to $0$ or $1$ which are
necessary for fulfilling the generalized Boolean constraints. This
will become more clear in in next subsection.

The result is quite interesting. Independently of the fraction of
canalizing Boolean functions, the PER cores for some external
conditions start to exist as soon as $\alpha$ exceeds 1/2. Note that
also the fraction of concerned variables does not depend on
$x$. According to the above discussion that the PER core for a BN
without canalizing functions is uniquely determined and of purely
topological nature, the same is true for the set of variables which
are contained at least in one PER core. This set does not depend on
the choice of functions, but only on the underlying graph structure.

\begin{figure}[htb]
\vspace{1.2cm}
\begin{center}
\includegraphics[width=0.75\columnwidth]{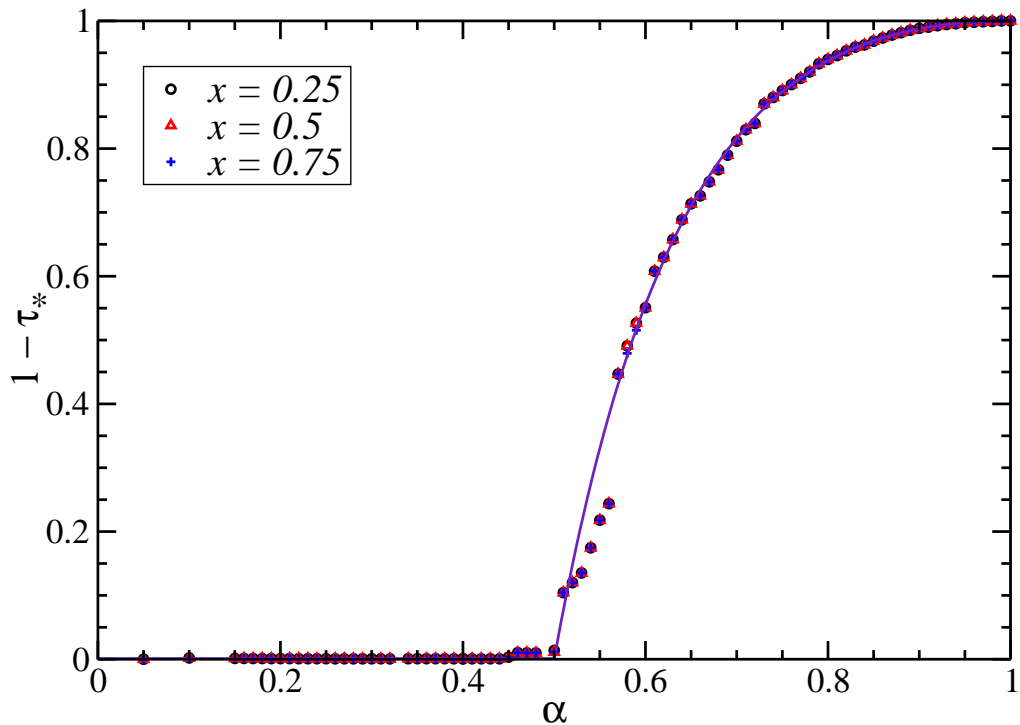}
\end{center}
\caption{Fraction of internal variables being in at least one PER
core: Analytical result (full line) versus BP results on single graphs
of $N=10000$ vertices.}
\label{fig:appearance}
\end{figure}

In Fig.~\ref{fig:appearance} we compare the analytical result to the
BP result on single BNs of $N=10000$ vertices, for various values of
$\alpha$. We plot the fraction $1-\tau_\star$ of all those internal
vertices which belong to the PER core for at least one external
configuration. Besides the good coincidence between analytical results
(derived in the thermodynamic limit) and single-sample data, one point
is remarkable: The BP curves for various $x$-values are practically
identical. They come from identical graphs with different realizations
of the functions, showing thus that the union of the PER cores for
all external conditions is a topological object of the underlying
graph.

\subsubsection{Dynamics}

To close this subsection, we discuss shortly the dynamics of
$\tau_*(t)$ as a time dependent quantity. Let us assume a random
asynchronous update. Then the expected number of sites contributing 
to $\tau_\star$ after $T$ single-variable updates is given by
\begin{equation}
M \tau_\star(T+1) = M \tau_\star(T)  -\tau_\star(T) 
+ (1-\alpha+\alpha\tau_\star(T))^2\ .
\end{equation}
On the right-hand side, we have three contributions: the number of
contributing sites after step $T$, the probability that the randomly
selected internal site already contributes, and the probability that
the inputs to this site at step $T$ force the variable to contribute
at step $T+1$. Note that we have already canceled the constantly
vanishing $\pi_0$ contribution in Eq.~(\ref{eq:tau_star0}). In the
thermodynamic limit we rescale the time as $t=T/M$, and the above
equation becomes an ordinary differential equation,
\begin{eqnarray}
\dot \tau_\star &=& -\tau_\star+(1-\alpha+\alpha\tau_\star)^2
\nonumber\\
&=& \left(\alpha^2\tau_\star -[1-\alpha]^2\right)\ (\tau_\star-1)
\end{eqnarray}
with the initial condition $\tau_\star(t=0)=0$. This equation is
solved by
\begin{equation}
\tau_\star(t) = \left\{
\begin{array}{cl}
\frac{(1-\alpha)^2}{\alpha^2} \frac {1-\exp\{(1-2\alpha)t\}}
{1- \frac{(1-\alpha)^2}{\alpha^2} \exp\{(1-2\alpha)t\}} 
& {\rm if}\ \alpha\neq\frac 12 \\
\frac t{t+4} & {\rm if}\ \alpha = \frac 12
\end{array}
\right.
\end{equation}
which, starting from zero, grows monotonously toward the fixed point
given in Eq.~\ref{eq:tau_star}. Note that the approach to the fixed
point is in general exponentially fast, but it slows down
algebraically at the critical point, there the asymptotic approach
happens as $t^{-1}$. Note also that, due to the initial condition
$\tau_\star(t=0)=0$ the dynamics stops as soon as it reaches the
smallest fixed-point of $\tau_\star$, which corresponds to the
selection of the largest number of $\star$, and justifies our previous
selection of the relevant solution.

\subsection{Appearance and size of the typical PER core}

\subsubsection{Fixed-point equations}

In the last subsection, we have seen that the first extensive PER
cores appear continuously at $\alpha=1/2$, independently of the value
of the fraction $x$ of non-canalizing functions. These cores
correspond, however, to an exponentially small fraction of all
external conditions. The average core size remains zero in the
thermodynamic limit. We can use the self-consistent equation
(\ref{eq:popdyn}) in order to determine the average fraction of
regulated nodes in the PER core,
\begin{equation}
n_{PER} = \int {\cal D} p \ {\cal P}[p] \ p(\star)\ .
\end{equation}
It can also be interpreted as the {\it typical} PER-core size, since a
randomly chosen external condition leads to this size with a
probability approaching one in the thermodynamic limit $N\to\infty$.

As already derived within a slightly different way in
\cite{letter,jstat} this size is given by the self-consistent equation
\begin{equation}
1-n_{PER} = ( 1 - \alpha n_{PER} )^2 + (1-x) ( 1 - \alpha
n_{PER} ) \alpha n_{PER}\ .
\end{equation}
The first contribution comes from a situation where both parent nodes
are not in the core, the second from canalizing functions with one
canalizing and one core input. This equation has two solutions, the
physically relevant one is the larger one:
\begin{equation}
n_{PER} = \left\{
\begin{array}{cl}
0 & {\rm if}\ \alpha\leq\frac 1{1+x} \\
\frac{\alpha(1+x)-1}{x\alpha^2}& {\rm if}\ \alpha>\frac 1{1+x}
\end{array}
\right.
\end{equation}
This implies that the PER core typically appears only at
$\alpha>1/(1+x)$, which goes from $\alpha=1$ at $x=0$ to $\alpha=1/2$ at
$x=1$. For purely canalizing networks we thus find that the typical
core size remains always zero, even if some rare core exist beyond
$\alpha=1/2$, whereas for purely non-canalizing BNs both threshold
coincide since the PER core in this case does not depend on the
external condition.

\subsubsection{Dynamics}

The dynamical evolution of this quantity can be derived following the
same procedure as in the last subsection. We immediately give the
resulting rate equation
\begin{equation}
\dot n_{PER}(t)=[\alpha(1+x)-1]\ n_{PER}(t) - \alpha^2\, x\, n_{PER}^2(t)
\end{equation}
Initially all internal variables are $\star$, so we have the initial
condition $n_{PER}(0)=1$.  This ordinary differential equation is
readily solved by
\begin{equation}
n_{PER}(t) = \left\{
\begin{array}{cl}
{\rm max}\left[ 0, g(t|\alpha,x)\right] &
{\rm if}\ \alpha\neq\frac 1{1+x} \\
\frac 1{1+\alpha^2\, x\, t}& {\rm if}\ \alpha=\frac 1{1+x}
\end{array}
\right.
\end{equation}
where 
\begin{equation}
g(t|\alpha,x) = \frac{(\alpha(1+x)-1)\exp\{[\alpha(1+x)-1]\, t\}  }{
x\alpha^2+[\alpha(1+x)-1-x\alpha^2]} 
\end{equation}

Note that for the subcritical case, the system reaches $n_{PER}=0$
after finite time, whereas it reaches its asymptotic value exponentially
fast in the supercritical phase. Exactly at the transition we see the
expected algebraic decay. 

\subsection{The intersection of all PER cores}

As a last special case which can be investigated analytically we look
at the intersection of the PER cores for all external conditions. The
question is: Are there variables never fixed by PER? How many of these
variables exist in a network? As before, we will first discuss the
fixed point equations for the size of the intersection, and derive the
phase-transition line for a non-trivial size. Than we are going to
discuss the dynamics of PER, and how the physically relevant solution
is selected.

\subsubsection{Fixed-point equations}

We are interested in the variables which are $\star$ for {\it all}
external conditions. Their fraction amongst the regulated variables
equals $\pi_\star$, according to Eq.~(\ref{eq:tau_pi}). Let us discuss
under which input conditions the output of a generalized Boolean
function becomes constantly $\star$:
\begin{itemize}
\item {\it Non-canalizing functions:} The output is fixed to $\star$
if and only if at least one input is regulated and fixed to
$\star$. This happens with probability $[1-(1-\alpha\pi_\star)^2]$.
\item {\it Canalizing functions:} Here the situation is a bit more
involved. The output is obviously $\star$ if both inputs are
$\star$. If only one input is $\star$, the other one is not allowed to
take its canalizing value - {\it i.e.} one input is ``frozen'' to
$\star$, the other ``frozen away'' from its canalizing value. The
cumulative probability of all these cases is $[2\alpha^2\pi_\star
\tau_0 - \alpha^2 \pi_\star^2]$. The negative term removes the
double-counting of the case of two $\star$-valued inputs. Note that
this is strictly true for the AND function having canalizing values
equal to zero for both inputs. For the other functions it follows from
the symmetry condition $\tau_0=\tau_1$.
\end{itemize}
We thus find
\begin{equation}
\label{eq:pi_star}
\pi_\star = x \left[1-(1-\alpha\pi_\star)^2\right] +
(1-x) \left[2\alpha^2\pi_\star \tau_0 - \alpha^2 \pi_\star^2\right]
\end{equation}
This equation depends still on $\tau_0$, {\it i.e.}, on the
probability that a variable is frozen away from $x=0$. We consider the
following cases
\begin{itemize}
\item {\it Non-canalizing functions:} There are two types of
contributions: In the first one, at least one input is frozen to
$\star$, {\it i.e.} the output becomes $\star$ too, and thus differs
from zero. The second contribution stems from the situation, where non
of the variables is frozen to $\star$, we have to avoid any input
configuration producing zero (00 and 11 for XOR, 01 and 10 for its
negation). The total probability for this is
$[1-(1-\alpha\pi_\star)^2 + 2\alpha^2(\tau_0-\pi_\star)^2]$.
\item {\it Canalizing functions, canalized output equals zero}: None
of the inputs is allowed to assume the canalizing value, which has
probability $\alpha^2\tau_0^2$.
\item {\it Canalizing functions, canalized output equals one}: Only
the simultaneous appearance of two non-canalizing input variables has
to be avoided, the corresponding probability reads
$[1-(1-\alpha\tau_0)^2]$.
\end{itemize}
The total equation for $\tau_0$ thus reads
\begin{equation}
\label{eq:tau_0}
\tau_0 = x\left[1-(1-\alpha\pi_\star)^2 
+ 2\alpha^2(\tau_0-\pi_\star)^2\right] + (1-x)\alpha\tau_0\ .
\end{equation}
Eqs.~(\ref{eq:pi_star},\ref{eq:tau_0}) are two closed quadratic
equations, thus they can be solved analytically. The solutions are:
\begin{enumerate}
\item The trivial solution
  \begin{equation}
    \pi_\star = \tau_0 = 0
  \end{equation}
\item A second solution of trivial $\pi_\star$
  \begin{equation}
    \pi_\star = 0,\ \ \ \ \ \tau_0 = \frac{1+\alpha(x-1)}{2\alpha^2 x}
  \end{equation}
\item The potentially physical solution
  \begin{eqnarray}
    \pi_\star &=& \frac{ 1+x-4 x^2 +\alpha(-1-2x+3x^2+4x^3) +(x-1) D
    }{2\alpha^2 x(2x^2-1)}
    \nonumber\\
    \tau_0 &=& \frac{ 1-4 x^2 +\alpha(-1-3x+4x^2+8x^3) + D
    }{4\alpha^2 x(2x^2-1)}
  \end{eqnarray}
  with
  \begin{equation}
   D =  \sqrt{1+\alpha(-2-6x+8x^3)+\alpha^2(1+6x+x^2-8x^3)}
  \end{equation}
\item The unphysical solution
  \begin{eqnarray}
    \pi_\star &=& \frac{ 1+x-4 x^2 +\alpha(-1-2x+3x^2+4x^3) -(x-1) D
    }{2\alpha^2 x(2x^2-1)}
    \nonumber\\
    \tau_0 &=& \frac{ 1-4 x^2 +\alpha(-1-3x+4x^2+8x^3) - D
    }{4\alpha^2 x(2x^2-1)}
  \end{eqnarray}
\end{enumerate}

\begin{figure}[htb]
\vspace{1.5cm}
\begin{center}
\includegraphics[width=0.75\columnwidth]{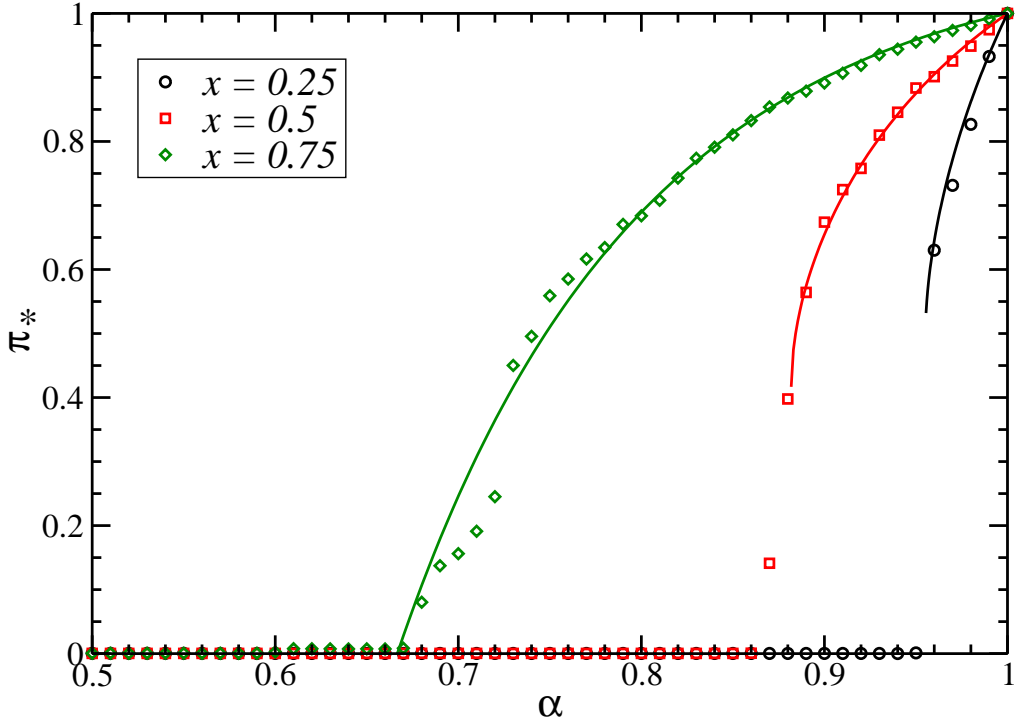}
\end{center}
\caption{Fraction of internal variables being in all PER cores:
Analytical result (full lines) versus BP results (symbols) on single
graphs of $N=10000$ vertices.}
\label{fig:intersection}
\end{figure}

The first solution is the correct one for small enough $\alpha$. The
third one becomes potentially relevant, when the argument under the
square root becomes positive, {\it i.e.}, when $D$ is a positive real
number. This point is actually relevant for small
$x$, for larger $x$ the corresponding $\pi_\star$ would be
negative. It becomes positive only at $\alpha=1/(2x)$ as can be seen
easily by linearizing the above equations. To summarize, the
non-trivial solution becomes physical beyond the phase transition line
\begin{equation}
\alpha(x) = \left\{
\begin{array}{cll}
\frac{1+3x-4x^3+2\sqrt{2x^2-6x^3+4x^6}}{1+6x+x^2-8x^3} &
{\rm for}\ x<\frac{1+\sqrt{17}}8 &\\
\frac1{2x} &{\rm for}\ x\geq\frac{1+\sqrt{17}}8 &\\
\end{array}
\right.
\end{equation}
In this equation, we have already indicated the character of the
transition: For $x$ below the tricritical point $(1+\sqrt{17})/8
\simeq 0.640388 $, the transition is discontinuous, and the
intersection of all PER cores jumps from an empty set to a non-zero
fraction of all internal vertices. Above this point, a non-trivial
intersection appears continuously crossing the critical line. In
Fig.~\ref{fig:intersection}, this analytical result is shown to be
well-confirmed by the BP results on a single graph of 10000 vertices.

\subsubsection{Dynamics}

\begin{figure}[htb]
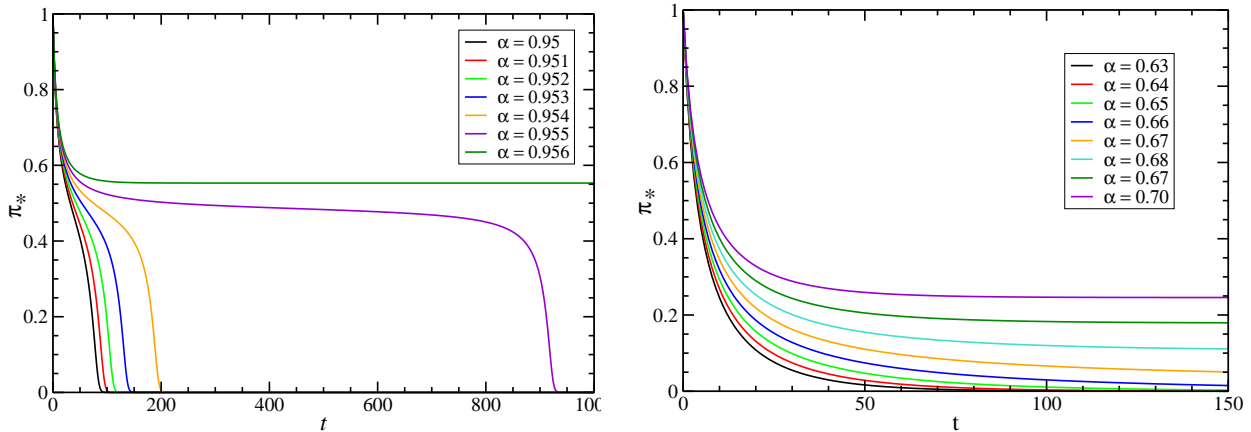

\vspace{1cm}
\begin{center}
\includegraphics[width=0.45\columnwidth]{intsec_syn_x025}\hspace{.2cm}
\includegraphics[width=0.45\columnwidth]{intsec_syn_x075}
\end{center}
\caption{Time dependence of the number of variables frozen to $\star$
under random asynchronous PER, for $x=0.25$ (left figure) and $x=0.75$
(right figure). For small $x$, the discontinuous character of the
transition can be beautifully seen from the appearance of a non-zero
plateau, whose length diverges at the phase-transition point
($\alpha\simeq 0.95505$ for $x=0.25$). The transition is continuous for
large $x$, as can be seen in the right figure.}
\label{fig:intsec}
\end{figure}

\begin{figure}[htb]
\vspace{1cm}
\begin{center}
\includegraphics[width=0.75\columnwidth]{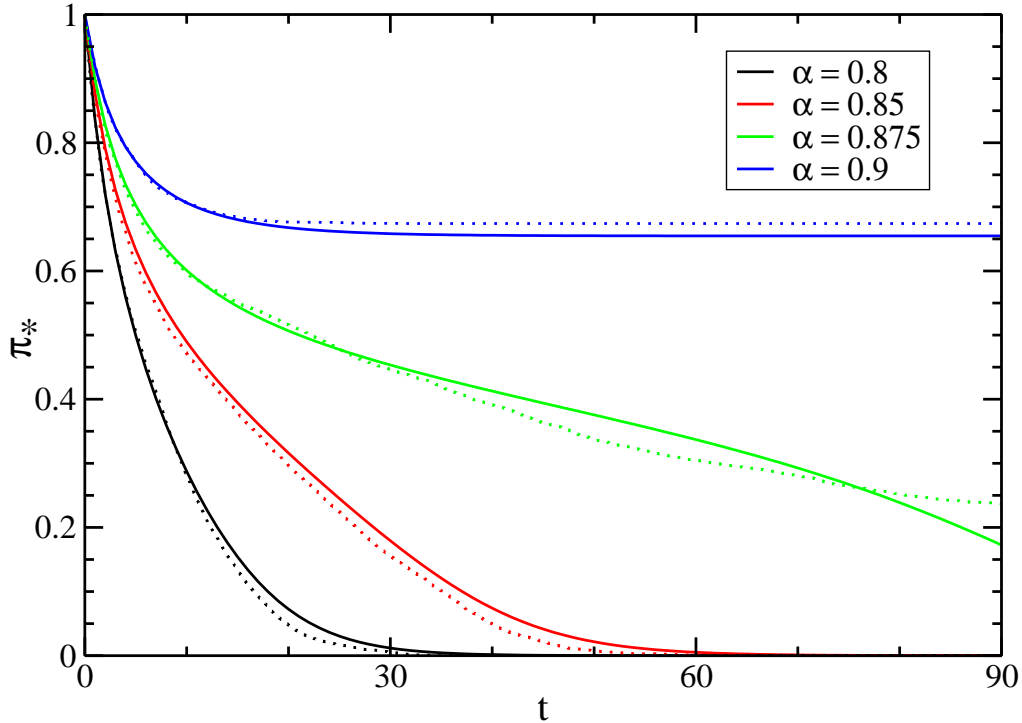}
\end{center}
\caption{Comparison of the analytical description (full lines) with
realizations in single random BNs of $N=10000$ vertices (dotted
lines). Up to sample-to-sample fluctuations, the results coincide.}
\label{fig:intsec2}
\end{figure}

As before, the selection of the relevant solution can be justified by
the dynamical behavior of PER. Following the same arguments as above,
we can write down two coupled ordinary differential equations for the
evolution of both $\pi_\star$ and $\tau_0$ which are expected to be
exact in the thermodynamic limit of large BNs. The equations are
\begin{eqnarray}
\label{eq:pi_star_dyn}
\dot\pi_\star 
&=& - \pi_\star + x \left[1-(1-\alpha\pi_\star)^2\right] +\nonumber\\
&&(1-x) \left[2\alpha^2\pi_\star \tau_0 - \alpha^2 \pi_\star^2\right]
\nonumber\\
\dot\tau_0
&=& -\tau_0 + x\left[1-(1-\alpha\pi_\star)^2 
+ 2\alpha^2(\tau_0-\pi_\star)^2\right] +\nonumber\\ 
&&(1-x)\alpha\tau_0
\end{eqnarray}
which have to be solved simultaneously under the initial condition
\begin{equation}
\pi_\star(t=0)=\tau_0(t=0)=1
\end{equation}
corresponding to the fact that initially all internal variables are
set constantly to $\star$. Unfortunately, we did not find a closed
analytical form for the solution of Eqs.~(\ref{eq:pi_star_dyn}), the
results of a numerical integration are represented in
Fig.~\ref{fig:intsec}. One can beautifully see the difference between
the discontinuous transition at low $x$ via the formation of a
finite-hight plateau, and the continuous transition at higher fraction
$x$ of non-canalizing functions. In Fig.~\ref{fig:intsec2} the result
of Eqs.~(\ref{eq:pi_star_dyn}) is plotted together with the actual BP
dynamics on a single BN of $N=10000$ variables, for $x=0.5$ and
various values of $\alpha$ below and above the transition point
$\alpha(x)\simeq 0.88185$. Both are completely consistent, as to be
expected sample-to-sample fluctuations are large close to the
transition, such that the self-averaging properties of $\pi_\star$ and
$\tau_0$ show up clearly only for larger samples.

\subsection{The phase diagram}
\begin{figure}[htb]
\vspace{1cm}
\begin{center}
\includegraphics[width=0.75\columnwidth]{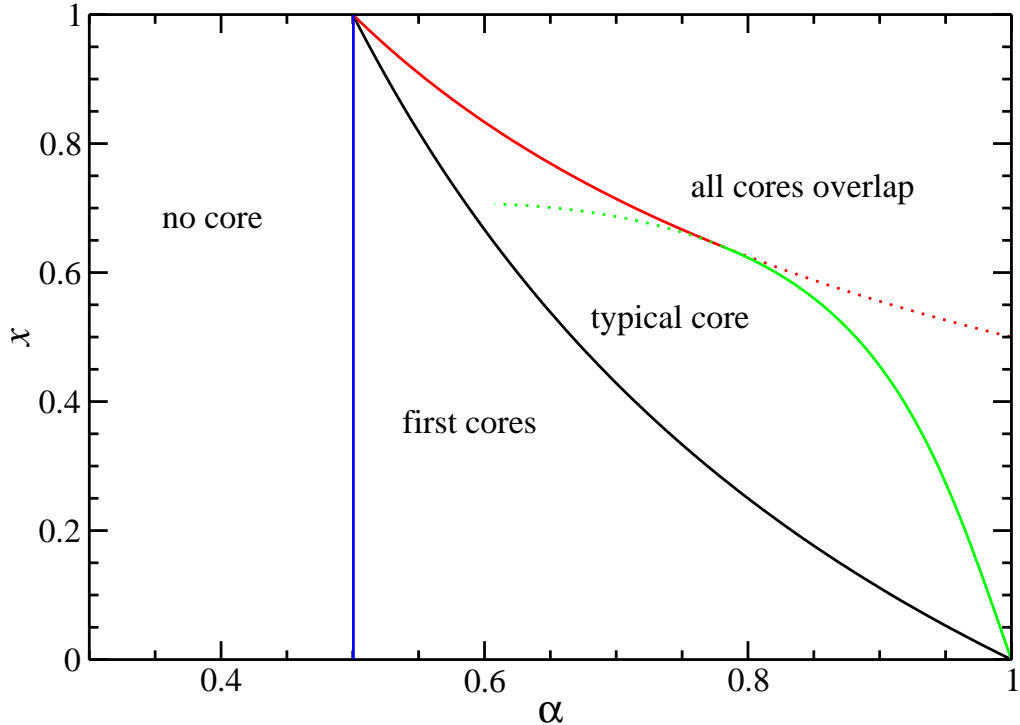}
\end{center}
\caption{Phase diagram for PER on random Boolean networks. The
dotted lines correspond to the unphysical parts of the phase
transition lines.}
\label{fig:phase_diagram}
\end{figure}

The findings are summarized in the phase diagram in
Fig.~\ref{fig:phase_diagram}. We can distinguish four different
regions
\begin{itemize}
\item For $\alpha<0.5$ (left of the blue vertical line), all external
information propagates throughout the system, with probability one no
PER core exists.
\item For $\alpha$ between the blue and the black lines, some external
conditions lead to a PER core. Almost all conditions, however,
propagate throughout the full systems, a random external condition
leaves almost surely no core.
\item For $\alpha$ in between the black and the red/green curves,
a typical external condition leads to an extensive core. The core size
starts to grow continuously at the black line, and increases when
entering deeper into this phase.
\item For $\alpha$ beyond the red/green curve, all PER cores overlap,
the intersection of all cores is non-empty. The transition is
discontinuous for small $x$ (green line), and continuous for larger
$x$ (red line).

\end{itemize}

\subsection{The distribution of PER-core sizes}

It is interesting to see directly what the distribution of the PER
cores size look like in single instances of BNs. To do so we have
generated graphs at $\alpha=0.95$ for different concentrations of
non-canalizing functions $x=(0 , 0.1, 0.4)$ and different sizes
$N=100,200,400$. Given a single instance of the BN one can calculate
the PER core sizes generated for each of the $2^{(1-\alpha)N}$
configurations of external variables, and finally compute their
distribution. Although the time needed for computing a PER core is
linear in $N$, we have to explore an exponential of number of external
conditions, which constrains us to work at relatively small $N$ and
at values of $\alpha$ very close to 1.

\begin{figure}[htb]
\vspace{0.1cm}
\begin{center}
\includegraphics[width=0.6\columnwidth]{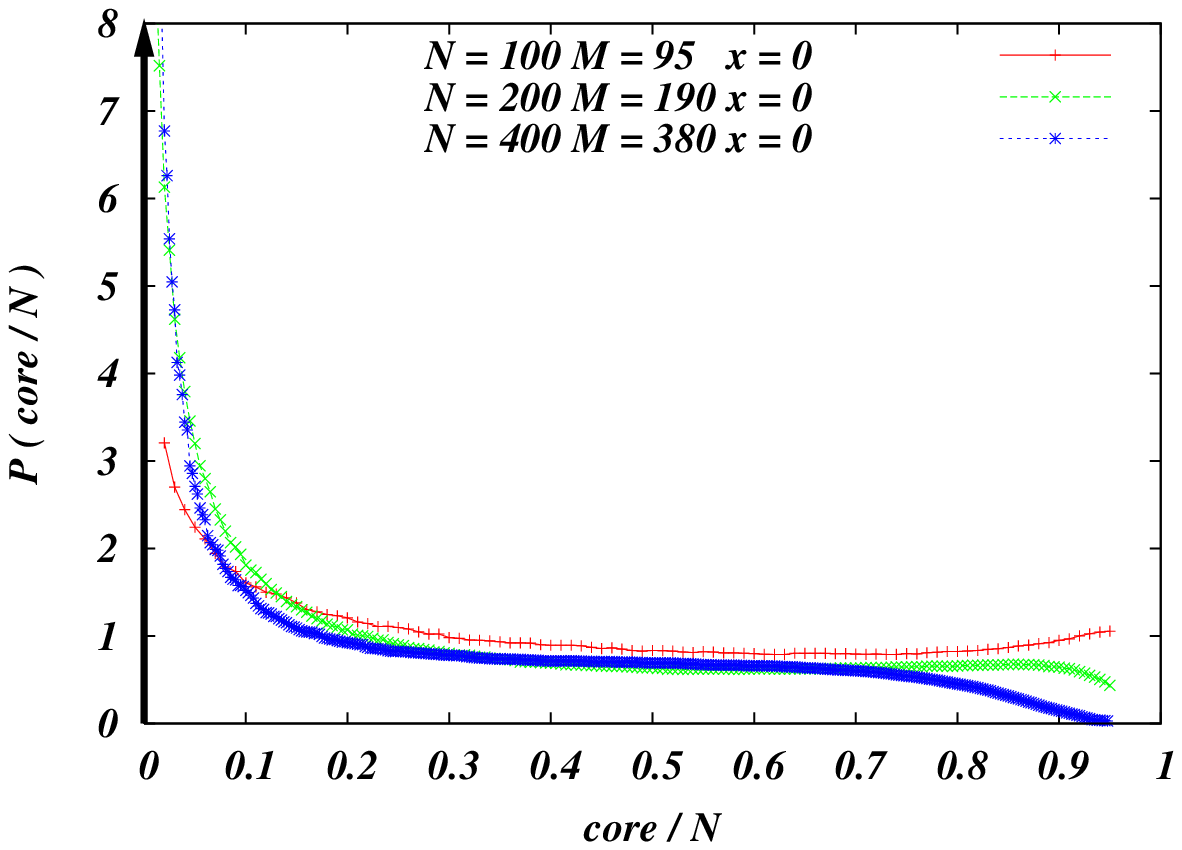}
\includegraphics[width=0.6\columnwidth]{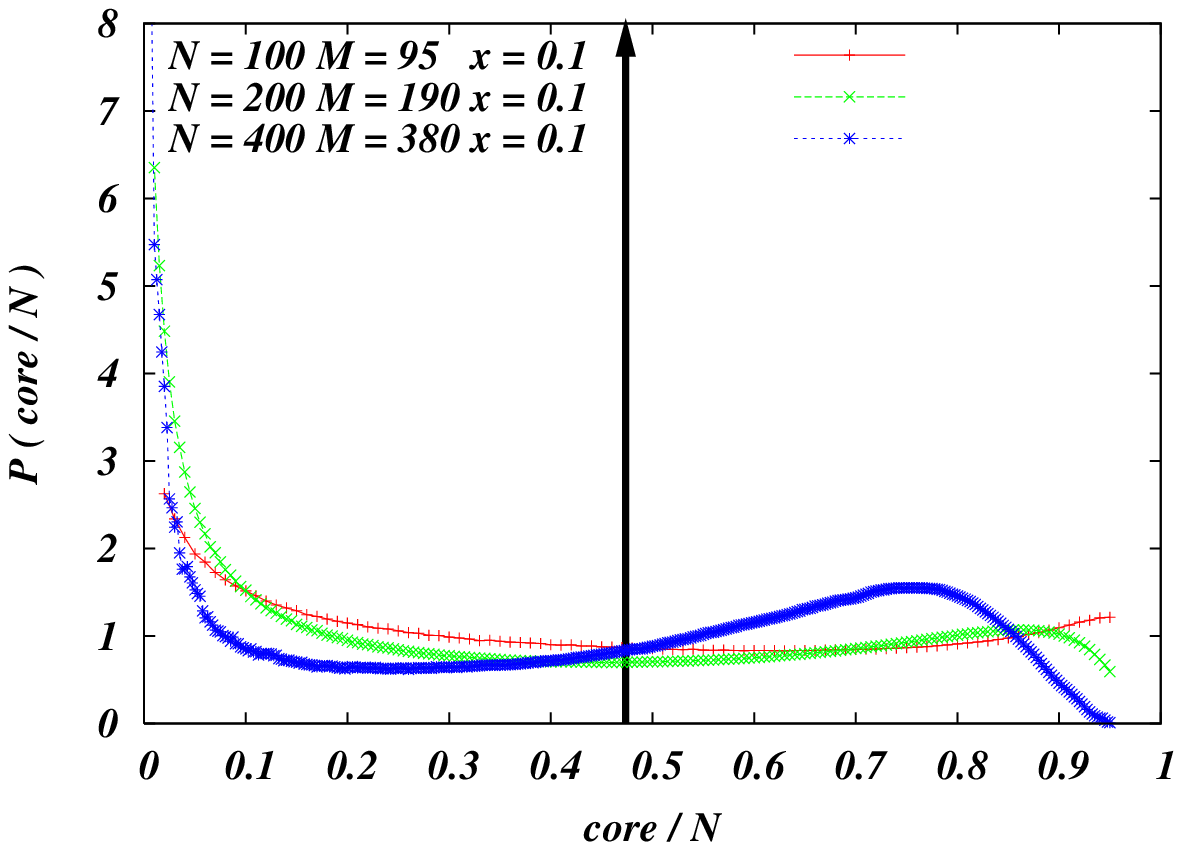}
\includegraphics[width=0.6\columnwidth]{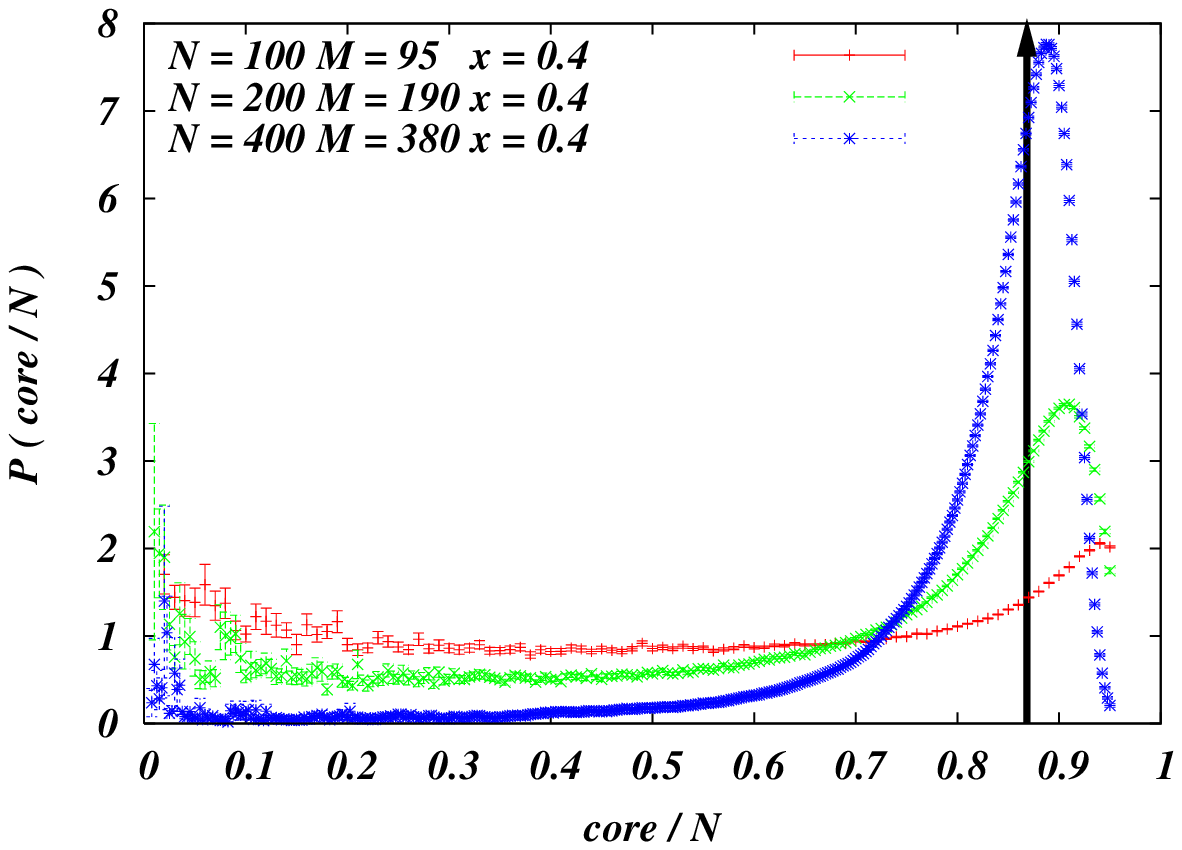}
\end{center}
\caption{Distribution of core sizes at $\alpha = 0.95$ for
two different concentration of non-canalizing functions $x=0.0
,0.1,0.4$ corresponding respectively to the {\em ``first cores''} phase
(left panel), the {\em ``typical core''} phase (central panel), and the
{\em ``all cores overlapping''} phase (right panel) respectively (see
Fig.~\ref{fig:phase_diagram}). The black solid arrows are the
theoretical average core density values in the thermodynamic limit for
the corresponding $(\alpha, x)$ values as computed in \cite{jstat}.}
\label{fig:core_histo}. 
\end{figure}

In Fig.~\ref{fig:core_histo} we display the histogram of
the the core size density averaged over 60000 different
samples. We also display the theoretical value for the average
core size as computed already in  \cite{jstat}
(displayed as a black thick arrows). The emerging scenario is
coherent with the analytic predictions:
\begin{itemize}
\item For $x=0$ we are in the {\em ``first cores''} phase where each of
  the external conditions fix almost all the variables as indicated by
  the concentration of the weight of the histogram around 0 (see left
  panel in Fig.~\ref{fig:core_histo}).
\item For $x=0.1$ we are in the {\em ``typical cores''} phase
  characterized by the formation of the small peak at high values of
  the core size density. Yet many of the external conditions fix
  almost all the variables as indicated by the concentration of the
  weight of the histogram around 0 (see left panel in
  Fig.~\ref{fig:core_histo}). In this region finite size scaling
  corrections are relevant probably due to the nearby phase transition
  line.
\item For $x=0.4$ we are in the {\em ``all cores overlapping''} phase and
  the histogram peaks nicely close to the theoretical mean value. The
  noisy signal at small core sizes is due to strong sample to sample
  fluctuations.
\end{itemize}

\section{On the random asynchronous dynamics of BN}
\label{sec:async_dyn}
In the previous sections, we have only analyzed the propagation of
external regulation. This last section is dedicated to the full
asynchronous update dynamics: Starting from a random assignment to all
variables, in every step one function is selected randomly and its
output is updated, {\it i.e.}, it changes its value if and only if the
function was unsatisfied before. As before, the time $t=T/M$ is
measured by rescaling the number of update attempts $T$ by the number
$M$ of all functions, such that in a time-interval of length one each
function is visited once on average. In this section we first give
an approximate description of the evolution of the number of
unsatisfied functions, and at the end we relate the dynamics to the
PER core studied before.

\subsection{The energy of canalizing and non-canalizing functions}
In order to characterize the asynchronous dynamics we introduce a
Hamiltonian which counts the number of unsatisfied Boolean
constraints:
\begin{eqnarray}
\label{eq:H}
{\cal H}(\vec s) &=&  \sum_{a\in INT} s_a \oplus
F_a(s_{a_1},...,s_{a_K}) \nonumber\\
&=& [ (1-x)\, e_c + x\, e_{nc} ] M \ .
\end{eqnarray}
The symbol $\oplus$ stands for the logical XOR operation, {\em
i.e.}~each cost term contributes $0$ to the sum if
Eq.~(\ref{eq:function}) is fulfilled, and $1$ otherwise. We have
divided this energy into contributions coming from canalizing
functions, and those from non-canalizing functions. More precisely,
$e_c$ stands for the fraction of unsatisfied canalizing functions,
$e_{nc}$ for the fraction of unsatisfied non-canalizing functions. In
the beginning, all variables take random values.
Each function is therefore violated with probability 1/2,
\begin{equation}
e_c(t=0) = e_{nc}(t=0) = \frac 12\ .
\end{equation}
Despite some efforts in the last years \cite{SeCu,DYN,SeWe,HanWe,Ha+},
the dynamics of diluted models remains an unsolved problem, and we
have to restrict our analysis to approximate methods. Following the
ideas of \cite{SeWe}, and restricting our attention to the simplest
level of description, we can close the equations for $e_c(t)$ and
$e_{nc}(t)$ by assuming that at each step all configurations of
given energy values are equally probable \cite{DRT}. This is clearly an
approximate assumption, but it leads to a good quantitative
description of the real dynamics.

As in the case of the core evolution, the dynamics leads to simple
ordinary differential equations. We first write them down, and then
discuss the meaning of all contributions:
\begin{widetext}
\begin{eqnarray}
\label{eq:e_approx}
(1-x)\, \dot e_c &=& - (1-x)\, e_c + [(1-x)\, e_c +x\, e_{nc}]\,
\alpha (1-x) (1-2e_{c}) \nonumber\\
x\, \dot e_{nc} &=& - x\, e_{nc} + [(1-x)\, e_c +x\, e_{nc}]\,
2\alpha x (1-2e_{nc})
\end{eqnarray}
\end{widetext}
Let us first discuss the first equation. The prefactor $(1-x)$ on the
left-hand side comes from the fact that the time is measured with
respect to $M$, whereas $e_c$ is a fraction of the $(1-x)M$ canalizing
functions. The first term on the right-hand side is the contribution
of the selected function itself: With probability $(1-x)$ it is
canalizing, and with probability $e_c$ it changes from unsatisfied to
satisfied, and the energy decreases by one. The second term comes from
the functions regulated by the output of the updated functions: With
probability $[(1-x)\, e_c +x\, e_{nc}]$ the selected function was
updated, and thus the dependent functions might change status. There
are on average $2\alpha$ such functions, a fraction $(1-x)$ of them
being canalizing. They change status only if the other input has not
its canalizing value (factor 1/2). The energy goes up for the fraction
$(1-e_c)$ of satisfied functions, and down for the fraction $e_c$ of
unsatisfied functions. The second equation is similar, with the main
difference that the energy status of a non-canalizing function changes
always if one input is changed, leading to a relative factor $2$
compared to the canalizing case.

These two equations have the obvious fixed point $e_c=e_{nc}=0$. It
corresponds to a fixed point of the microscopic dynamics in the sense
that all functions are satisfied, and the update dynamics never flips
any of the variables. Are there also other, positive-energy solutions?
Assuming that a non-zero solution appears continuously, we linearize
the fixed point equation
\begin{eqnarray}
0 &=& - (1-x)\, e_c + [(1-x)\, e_c +x\, e_{nc}]\,
\alpha (1-x)  \nonumber\\
0 &=& - x\, e_{nc} + [(1-x)\, e_c +x\, e_{nc}]\,
2\alpha x
\end{eqnarray}
Adding these two equations, we find that the non-zero solution appears
at
\begin{equation}
\alpha = \frac 1{1+x}\ ,
\end{equation}
{\it i.e.} at the same point where typically the PER core
appears. This transition corresponds to critical BN \cite{jstat}. 
For smaller $\alpha$, the system fastly approaches a zero-energy fixed 
point, the basic mechanism being fixation of almost all 
variables by PER. Above the transition, the system settles down at
positive energy. At least a part of the core variables continues to 
flip for ever (in the thermodynamic limit).

\begin{figure}[htb]
\vspace{1cm}
\begin{center}
\includegraphics[width=0.75\columnwidth]{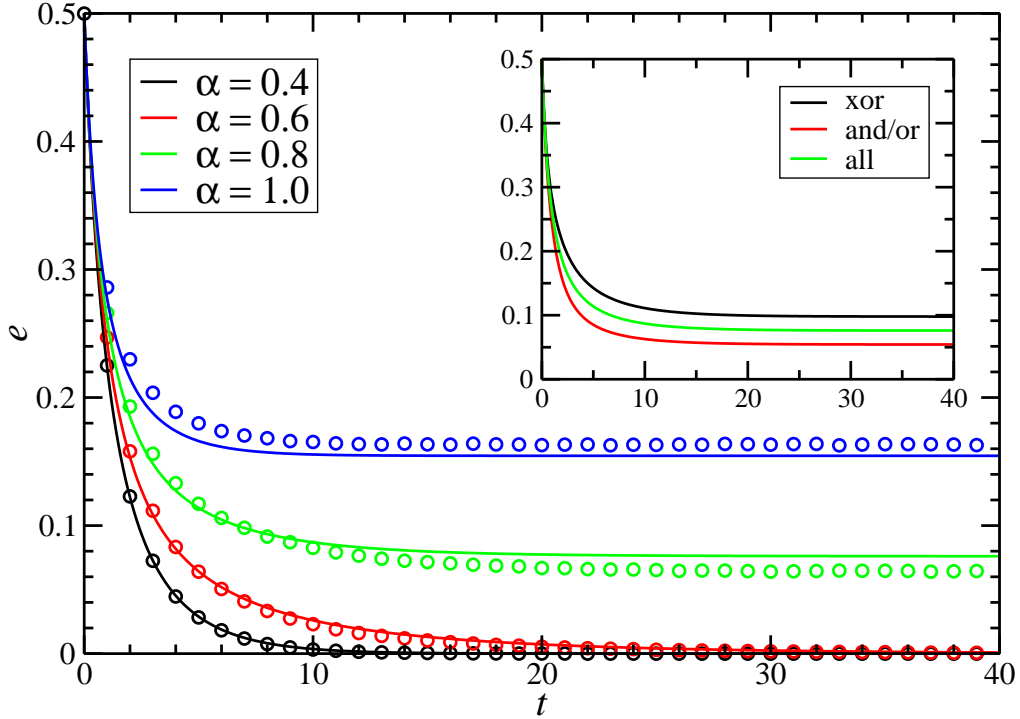}
\end{center}
\caption{Time evolution of the energy: Approximate analytical
description (full lines) vs. numerical simulations (symbols) on single
BNs of $N=10^6$. The inset shows the different evolution of the energy
contributions coming from canalizing and non-canalizing functions.}
\label{fig:energy}
\end{figure}

To confirm this picture, we have numerically integrated
Eqs.~(\ref{eq:e_approx}) and in parallel performed direct numerical
simulations on large BN. The results are shown in
Fig.~\ref{fig:energy}, and confirm the analytical findings. We see, in
particular, that the simple approximation already gives a
quantitatively good description of the dynamics. Following the ideas
of \cite{SeWe} this result can surely be improved, but doing so goes
beyond the scope of the present paper.

\subsection{Relating PER and dynamics}

In Sec.~\ref{sec:per} we have studied PER also as a dynamical process.
In the initial condition all external variables are fixed to some 
configuration in $\{0,1\}^{N-M}$, whereas internal variables are
initially assigned the joker value $\star$. This generalized configuration 
can be understood as a projection of the set of {\it all} $2^{\alpha N}$ 
initial conditions with fixed external and changeable internal nodes. 
This consideration makes clear that (for the fixed external configuration)
all variables {\it not} belonging to the PER core get fixed after 
${\cal O}(N \log N)$ elementary steps. Core variables potentially might 
go on flipping for ever. However, also subsets of core variables could 
freeze self-consistently due to feedback loops. These subsets depend
both on the initial condition of the regulated variables and on the
realization of random update dynamics.

\begin{figure}[htb]
\begin{center}
\includegraphics[width=0.75\columnwidth]{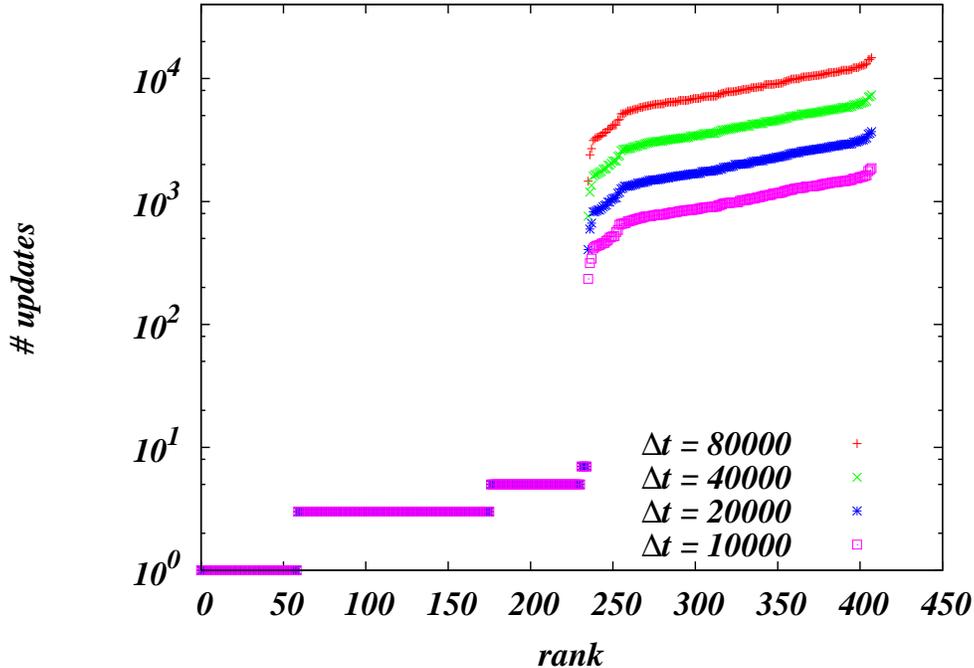}
\end{center}
\caption{Ranked distribution of the number of elementary spin-updates
of the set of variables belonging to the PER core, for $N=1000$, $M=900$, 
$x=0.2$ and up to 80000 asynchronous updates sweeps (see text). 
The PER core in this case is made of 408 variables. The 250 first-ranking 
spins fix their value after $\sim 10$ steps while the remaining spins keep
changing their values.}
\label{fig:dyncore}. 
\end{figure}

To investigate this question, we have analyzed the dynamics restricted
to the PER core. We first fixed all non-core variables by PER, note that
this can achieved efficiently for any external condition. Then we set 
the core variables
randomly. This slightly atypical choice of initial condition helps to
make more evident the dynamics on the core. In the asynchronous
dynamics, all these variables (or their regulating function) are
on average seen equally often, but they are flipped if and only if 
the function was not satisfied before. We record the number
of actual variable changes for each PER core site, in order to identify
frozen versus changeable variables.

In Fig.~\ref{fig:dyncore} we display the number of actual spin-flips for 
each variable during a run of 80000 sweeps (each sweep consisting of 
$\alpha N$ attempted spin flips). We measure these quantities at different 
times, namely after $t = 10000,20000,40000$ and $80000$ sweeps. We consider 
a random BN with $N=1000$, $M=\alpha N=900$, $x=0.2$. The configuration
of the external inputs is chosen at random and leads to a PER core of 408 
variables.

We find two types of behavior: In this sample more than half of the
variables become frozen at the very beginning of the dynamics and change 
only up to about $10$ times. This happens despite the fact that they
belong to the core and thus are not fixed by PER. Freezing thus has to
appear self-consistently due to feedback loops. The second group of
variables goes on changing. The number of their spin-flips grows
proportionally to the number of sweeps, as follows from the parallel 
curves in Fig.~\ref{fig:dyncore}. After a few sweeps the dynamics
becomes concentrated to this second class of variables. Note however
that the subdivision of core variables into these two classes depends
on the initial condition and the order of the updates which by 
definition is random. Both identity and number of frozen spins fluctuate
from realization to realization, but the qualitative scenario remains the
same. It would be very interesting to get a quantitative analytical 
understanding of this phenomenon.

\section{Conclusions}
\label{sec:concl}

In this work we have studied the propagation of external regulatory 
information into a random Boolean network. The efficiency of this 
propagation depends on two control parameters: The fraction 
$(1-\alpha)$ of non-regulated input variables in between all Boolean
variables, and the fraction $x$ of non-canalizing variables.

We find that for $\alpha<1/2$ the external condition propagates 
through the full BN and fixes efficiently all variables. For higher
$\alpha$, a fraction of variables remain unfixed by this process, forming
thus the PER core of the BN. The precise PER core depends on the external
condition. We have studied the fluctuation of the PER core size between
different external conditions. Only for $\alpha>1/(1+x)$ the PER core is 
not empty for almost all external conditions, for even higher $\alpha$
all PER cores start to overlap.

The notion of the PER core is intimately related to the dynamics of the
system. After very short time all non-core variables become frozen, so
only core variables can be considered as the true dynamical degrees of
freedom of the BN under a specified external configuration. However, also
a part of the core variables becomes dynamically frozen during the
dynamical evolution of the system, the reason being self-consistently
fixed feedback loops. The dynamics of the system for longer time scales
becomes concentrated on the remaining variables.

In a future work, we plan to extend the PER analysis also to the
analytical calculation of the PER core size distribution, which so far
we obtained only numerically for small BNs. In this context it would
be interesting to algorithmically identify those external conditions
which lead to largest or smallest cores. This would help to solve the 
{\it inverse} problem of determining the external conditions leading to 
a specific core.

A second interesting extension of the current work would be a deepened
analysis of the dynamical behavior of this model. In distinction to
other models on diluted networks, including in particular diluted spin
glasses, BN are defined on {\it directed} graphs. This leads to some
technical simplifications which we hope to open new ways in the
understanding of their dynamical behavior.

\end{document}